\documentclass[english,12pt,twoside]{article}
\usepackage{amssymb}
\usepackage{amsmath}
\usepackage{babel}
\usepackage{array}
\usepackage[dvips]{graphicx}
\usepackage{pstricks}
\usepackage{pst-node}
\begin{document}
\date{}
\title{Quantum Cat's Dilemma}
\markboth{Marcin Makowski and~Edward W.~Piotrowski}{Quantum cat's dilemma}
\author{Marcin Makowski\footnote{mmakowski@alpha.pl} \,and~Edward W.~Piotrowski\footnote{ep@alpha.uwb.edu.pl}\\[1ex]\small
Institute of Mathematics, University of Bia\l ystok,\\\small
 Lipowa 41, Pl 15424 Bia\l ystok,
Poland}
\maketitle
\begin{abstract}
  We study a quantum version of the sequential game illustrating problems connected with making rational decisions. We compare the results that the two models (quantum and classical) yield.  In the quantum model intransitivity  gains importance significantly. We argue that the quantum model  describes our spontaneously shown preferences more precisely than the classical model, as these preferences are often intransitive.
\end{abstract}

Keywords: quantum strategies; quantum games; intransitivity; sequential game.

\section{Introduction}
A fundamental scientific theory is marked by its ability to solve
the widest possible range of problems. In the 20th century, it was
quantum mechanics \cite{r2} that became such an effective panacea
for the problems that could not be either understood or solved
with the use of the traditional methods. Quantum mechanics
describes the  fascinating  structure  of
elements of which the world is composed and explains such
phenomena as radioactivity, antimatter, stability of  molecular
structures, stars evolution etc. Quantum theory  allows
us to abandon the traditional paradigms of perceiving the world. Quantum-like  ideas are used in various fields of research and in this way they contribute to the unification of modern
science. Some of the mechanisms characteristic for living nature
may find their reflection in quantum theory \cite{r4}. Presently,
quantum information theory is being built at the meeting point of
quantum mechanics and theory of information \cite{r5,r6,r7}. The
concept of a quantum computer  stress  the
qualitative limitations of orthodox Turing machines
which in the future  would probably be replaced with the
quantum computers whose counting ability will substantially exceed
the possibilities of the  present computers \cite{r8}. It
poses a threat of using quantum technology to jeopardize the
contemporary methods used to guarantee the confidentiality of data
transfer \cite{r32}. It seems that the methods of quantum
cryptography that are being presently worked out will remain safe
even in the times of the quantum computers \cite{r9}. The
combination of the research methods of both information and game
theories results in emerging of the new mysterious field - quantum
game theory, in which the subtle quantum rules characterizing the
material world determine ways of controlling and
transformation of information \cite{r10, r11, r12, r13, r14}. In the quantum game formalism, pure strategies correspond to
the vectors of Hilbert space (to be more precise: the projective
operators on  subspaces determined by these vectors). The
mixed strategies are represented by the convex combinations of
vectors projected on these directions. In comparison with the sets
of the traditional strategies,  quantum strategies provide
players with much more possibilities which they can use while
making the most beneficial decision for themselves. This
characteristic feature of quantum game theory is the reason why
its results go beyond the traditional boundaries \cite{r33}.
Plenty of quantum variants of problems  analysed by the
traditional classical game theory (see \cite{r15, r30, 14}) 
have already been put forward.  First attempts at creating
quantum economy by applying quantum game theory to selected
economic problems have been made too  \cite{r16}. It is assumed
that there exists a market where financial transactions are made
with help of quantum computers operating on quantum strategies
\cite{r16, r17, 16}. It is essential to mention here that game
theory in its traditional  form has been formulated in the
context of  economic issues.\newline  The quantum game 
formalism has already been used to describe the idea of the
Evolutionary Stable Strategy(ESS) \cite{r31}. Perhaps, further
research in this direction will be used to explain a number of
phenomena that are now being researched by evolutionary biology.

In our work we  concentrate on the quantitative analysis of the
quantum version of a very simple game against Nature which was
presented and analyzed in  \cite{r1}. To illustrate the problem,
we will use the story about Pitts's experiments with cats,
mentioned in the Steinhaus diary \cite{r21}. Let us assume (alike
as in \cite{r1}), that a cat (we will be calling it the {\em
quantum cat}\/) is offered three types of food (no. 1, no. 2 and
no. 3), every time in pairs of two types, whereas the food
portions are equally attractive regarding the calories, and each
one has unique components that are necessary for the cat's good
health. Let us assume that the quantum cat reaches its optimal
strategy in  conditions of a  constant frequency of
appearance of a particular pairs of food in a diet, and that it
will never refrain from making the choice. The ability of finding
the optimal strategy by quantum cat may explain the principle of
last action formulated by Ernest Mach \cite{r20}. It is possible
that some of our psychological processes are subject to some
variant of the principle linked with Ockham's razor.

 Non-orthodox quantum description of the decision algorithms provides a possibility to extend  the results of Ref. \cite{r1}. In the following paragraphs, we  compare the quantum and the classical variants of the model we are interested in.
\section{Intransitivity}
However, before we start analyzing all possible behavioral patterns of quantum cat, it would be advisable to explain what the intransitive order is.
    Any relation  $\succ$ existing between the elements of a certain set is called \emph{transitive}\/ if $A\succ C$ results from the fact that $A\succ B$ and $B\succ C$ for any three elements $A$, $B$, $C$. If this condition is not fulfilled then the relation will be called \emph{intransitive}\/.

The best known example of intransitivity is the children game
"Rock, Scissors, Paper" (see quantum analysis of this game
\cite{r19}). Another interesting example of intransitive
order is Condorcet's voting paradox. Consideration regarding this
paradox led Arrow in the XX-th century to prove the theorem
stating that there is no procedure of successful choice that would
meet the democratic assumption \cite{r26} (some other problems
with intransitive options can be found in \cite{r25,r27}).
Intransitive orders still are surprisingly suspicious for many
researchers. Economists have long presented a view that people
should choose between things they like in a specific, linear order
\cite{r28}. But what we  actually prefer often depends on how
the choice is being offered \cite{r22,r23}. Mentioned in
Steinhaus's diary Pitts notice that a cat facing choice between
fish, meat and milk prefers fish to meat, meat to milk, and milk
to fish! Pitts's cat, thanks to the above-mentioned food
preferences, provided itself with a balanced diet.

Let us have a closer look at the effects of the consideration of the problem that Pitts's was trying to tackle, in the language of quantum game theory.

\section{Properties of cat's optimal strategies}
There is the following relation between the frequencies $\omega_k$\/, $k=0,1,2$ of appearance  of the particular foods in a diet and the conditional probabilities which we are interested in (\,see \cite{r1}):
\begin{equation}
\omega_k:=P(C_k)=\sum_{j=0}^{2}P(C_{k}|B_{j})P(B_{j}),\,\, k=0,1,2\,,
\end{equation}
where $P(C_{k} | B_{j})$ indicates the probability of choosing the food of number $k$\/, when the
offered food pair does not contain the food of number $j$\/, $P(B_{j})=:q_j$ indicates the frequency
of occurrence of pair of food that does not occur food number $j$\/.
The most valuable way of choosing the food by cat occurs for such six conditional probabilities
($P(C_1|B_0),$ $P(C_2|B_0)$,$
  P(C_0|B_1)$,$P(C_2|B_1)$,
  $P(C_0|B_2)$,
  $P(C_1|B_2)$)
  which fulfills the following condition:
\begin{equation}
 \label{maximum}
 \omega_0=\omega_1=\omega_2=\tfrac{1}{3}.
 \end{equation}
 Any six conditional probabilities, that for a fixed triple ($q_0,q_1,q_2$) fulfill (\ref{maximum})
 will be called a cat's \emph{optimal strategy}\/.
 The system of Eq. (\ref{maximum}) has the following matrix form:

  \begin{equation}\label{system}
\left( \begin{array}{ccc} P(C_0|B_2) & P(C_0|B_1) & 0 \\ P(C_1|B_2) & 0 &
P(C_1|B_0) \\ 0 & P(C_2|B_1) & P(C_2|B_0)
\end{array} \right)
\left( \begin{array}{ccc} q_2 \\ q_1 \\ q_0
\end{array} \right)=\tfrac{1}{3}
\left( \begin{array}{ccc} 1 \\ 1 \\ 1
\end{array} \right).
\end{equation}
and its solution:
 \begin{eqnarray}\label{solution}
q_2&=&\tfrac{1}{d}\bigg(\frac{P(C_0|B_1)+P(C_1|B_0)}{3}-P(C_0|B_1)P(C_1|B_0)\negthinspace\bigg),\nonumber\\
q_1&=&\tfrac{1}{d}\bigg(\frac{P(C_0|B_2)+P(C_2|B_0)}{3}-P(C_0|B_2)P(C_2|B_0)\negthinspace\bigg),\\
q_0&=&\tfrac{1}{d}\bigg(\frac{P(C_1|B_2)+P(C_2|B_1)}{3}-P(C_1|B_2)P(C_2|B_1)\negthinspace\bigg),\nonumber
 \end{eqnarray}
 defines a mapping $\mathcal{A}_0:D_3\rightarrow T_2$ of the three-dimensional cube ($D_3$\/)
 into a triangle ($T_2$\/)\,(two-dimensional simplex, $q_0+q_1+q_2=1$ and $q_i\geq 0$), where
 $d$ is the determinant of the matrix of parameters $P(C_j|B_k)$.
 The barycentric coordinates of a point of this triangle are interpreted as a probabilities
 $q_0, q_1$ and $q_2$.
 Thus we get relation between the optimal cat's strategy and frequencies $q_j$ of appearance
 of food pairs.

\section{Quantum cat}
We  start with the presentation of formalism which is indispensable for the quantum description of the variant of the game presented in the article \cite{r1}.
Let us denote tree different bases of two-dimensional Hilbert space as
 $\{\,| 1 \rangle\negthinspace_{\,\,0}, | 2 \rangle\negthinspace_{\,\,0}\,\}$,
$\{\,| 0 \rangle\negthinspace_{\,\,1}$, $| 2 \rangle\negthinspace_{\,\,1}\,\}$,
$\{\,| 0 \rangle\negthinspace_{\,\,2}$, $| 1 \rangle\negthinspace_{\,\,2}\,\}=\{\,
(1,0)^{T},(0,1)^{T}\,\}$.
The bases should be such that bases
\{\,$| 0 \rangle\negthinspace_{\,\,1}$, $| 2 \rangle\negthinspace_{\,\,1}$\,\},
\{\,$| 1 \rangle\negthinspace_{\,\,0}$, $| 2 \rangle\negthinspace_{\,\,0}$\,\}
are the image of
\{\,$| 0 \rangle\negthinspace_{\,\,2}$, $| 1 \rangle\negthinspace_{\,\,2}$\,\}
under the transformations $H$
 and $K$ respectively:\footnote{$H$ is called Hadamard matrix.}
\begin{displaymath}
\label{matrix maximal}
 H=\frac{1}{\sqrt{2}}\left(\begin{array}{cr}1 & 1 \\ 1 & -1
\end{array}\right)\/,\quad
 K=\frac{1}{\sqrt{2}}\left(\begin{array}{cr}1 & 1 \\ i & -i
\end{array}\right).
\end{displaymath}
It is worth to mention here that the set of so called conjugated bases, which is presented above,
allowed Wiesner (before asymmetric key cryptography was invented !)  to begin  research into  quantum cryptography.  These bases play also an important role in  universality of quantum market games \cite{1}.
Let us denote strategy of choosing the food number $k$\/, when the offered food pair not contain
the food of number $l$\/, as $| k \rangle\negthinspace_{\,\,l}$ ($k, l=0,1,2$, $k\ne l$).\newline
A family $\{|z\rangle\negthinspace\,\/\}$ (\,$z \in \overline{\mathbb{C}}$\,) of convex vectors:
\begin{displaymath}
 | z \rangle\negthinspace:=| 0 \rangle\negthinspace_{\,\,2}+z
|1\rangle\negthinspace_{\,\,2}=| 0 \rangle\negthinspace_{\,\,1}+\frac{1-z}{1+z}
|2\rangle\negthinspace_{\,\,1}=| 1 \rangle\negthinspace_{\,\,0}+\frac{1+iz}{1-iz}
|2\rangle\negthinspace_{\,\,0},
\end{displaymath}
defined by the parameters of the heterogeneous coordinates of the projective space $\mathbb{C}P^{1}$, represents all quantum cat strategies spanned  by the base vectors.
 The coordinates of the same strategy $| z \rangle\negthinspace$ read (measured) in various bases define quantum cat's preferences toward a food pair represented by the base vectors.
 Squares of their modules, after  normalization, measure the conditional probability of quantum cat's making decision in choosing a particular product, when the choice is related to the suggested food pair (the choice of the way of measuring a strategy).
 In this way, quantum cat makes a decision to choose the right food pair with the following probabilities:
 \begin{align}\label{derek}
 P(C_0|B_2) & =\frac{1}{1+|z|^{2}},&  P(C_1|B_2) &=\frac{|z|^{2}}{1+|z|^{2}}, \nonumber  \\
P(C_0|B_1)&=\frac{1}{1+|\frac{1-z}{1+z}|^2},
& P(C_2|B_1) &=
\frac{|\frac{1-z}{1+z}|^2}{1+|\frac{1-z}{1+z}|^2}, \\
P(C_1|B_0)&=\frac{1}{1+|\frac{1+iz}{1-iz}|^2},&
P(C_2|B_0)&= \frac{|\frac{1+iz}{1-iz}|^2}{1+|\frac{1+iz}{1-iz}|^2}\,.\nonumber
\end{align}
 Strategies $| z \rangle\negthinspace$\, can be parameterized by the sphere $S_2\backsimeq \overline{\mathbb{C}}$\, by using stereographic projection which establishes correspondence (bijection) between elements of $\overline{\mathbb{C}}$ and
 the points of $S_2$ (\,the north pole of the sphere corresponds with the point in infinity, $|\infty\rangle\negthinspace$ :=$| 1 \rangle\negthinspace_{\,\,2}$\,). Eq. $(\ref{derek})$ lead to the mapping $\mathcal{A}_1:S_2\rightarrow D_3$ of the strategies defined by the parameters of the sphere points onto the three-dimensional cube of conditional probabilities:
\begin{align}\label{prop}
P(C_0|B_2) &=\frac{1-x_3}{2}, & P(C_1|B_2) &=\frac{1+x_3}{2},\nonumber \\
P(C_0|B_1) &=\frac{1+x_1}{2},& P(C_2|B_1) &=\frac{1-x_1}{2}, \\
P(C_1|B_0) &=\frac{1+x_2}{2},& P(C_2|B_0) &=\frac{1-x_2}{2}\,.\nonumber
    \end{align}
     Combination  of the above projection with (\ref{solution}) results in the projection\newline $\mathcal{A}:S_2\rightarrow T_2$,    $\mathcal{A}:=\mathcal{A}_0\circ\mathcal{A}_1$
    of two-dimensional sphere $S_2$ into a triangle $T_2$.

The knowledge of  $\mathcal{A}$ allows to compare the number (measure) of the sets of the possible strategies of the quantum cat and the classical cat having the characteristics we are interested in.

\section{Quantum cat vs. Classical cat}
In this paragraph, we  compare the model described above, in which
quantum cat  can adopt strategies from any group of strategies
described above with the quantitative results of Ref. \cite{r1}.
In order to present the range of representation $\mathcal{A}$ of
our interest, we illustrated it with the values of this
representation for 10,000 randomly selected points with respect to
constant probability distribution on the sphere $S_2$\/. The
choice of such a measurement method for quantum cat's strategy
justifies the fact that that constant probability distribution
corresponds to the Fubini-Study measure on $\mathbb{C}P^{1}$
$\cite{r34}$ which is the only invariant measure in relation to
any change of quantum cat's decision regarding the chosen strategy
(the so called quantum tactic, homography on $\mathbb{C}P^{1}$).
Changes of the quantum cat's strategies therefore do not
influence the discussed below model.
\subsection{Optimal strategies}
\begin{figure}[htbp]
         \centering{
       \includegraphics[
          height=2in,
          width=2.1in]%
         {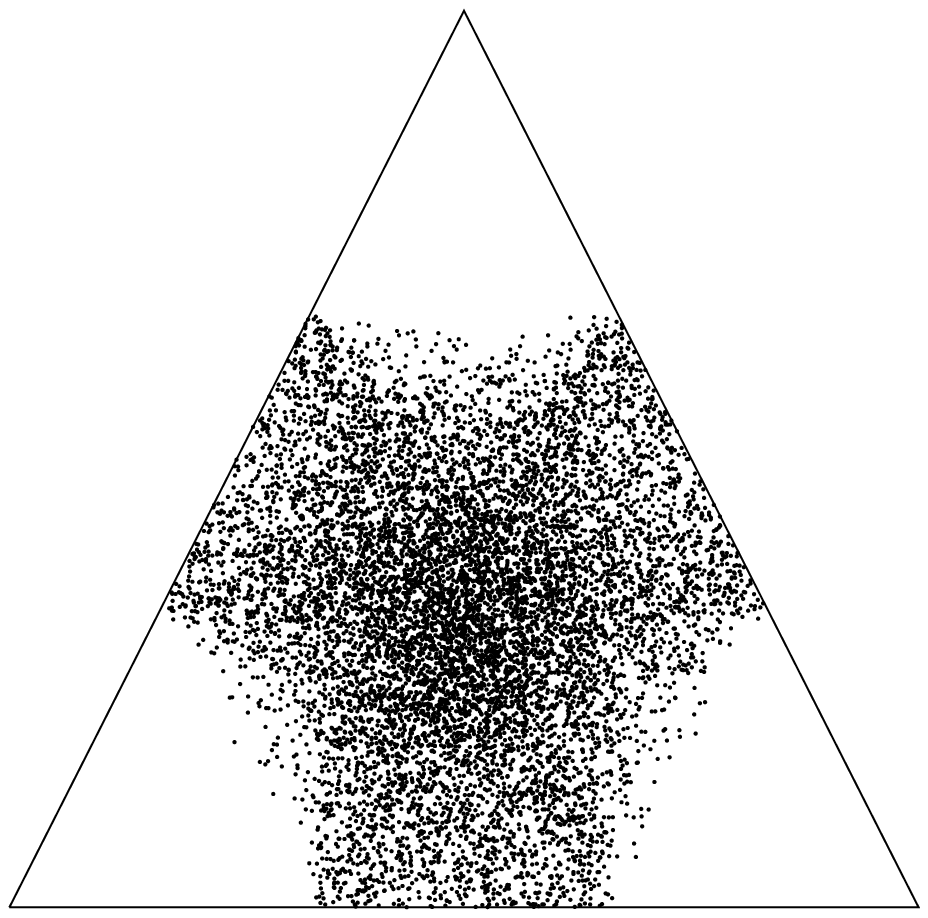}
         \includegraphics[
                      height=2in,
                      width=2.1
                      in]%
       {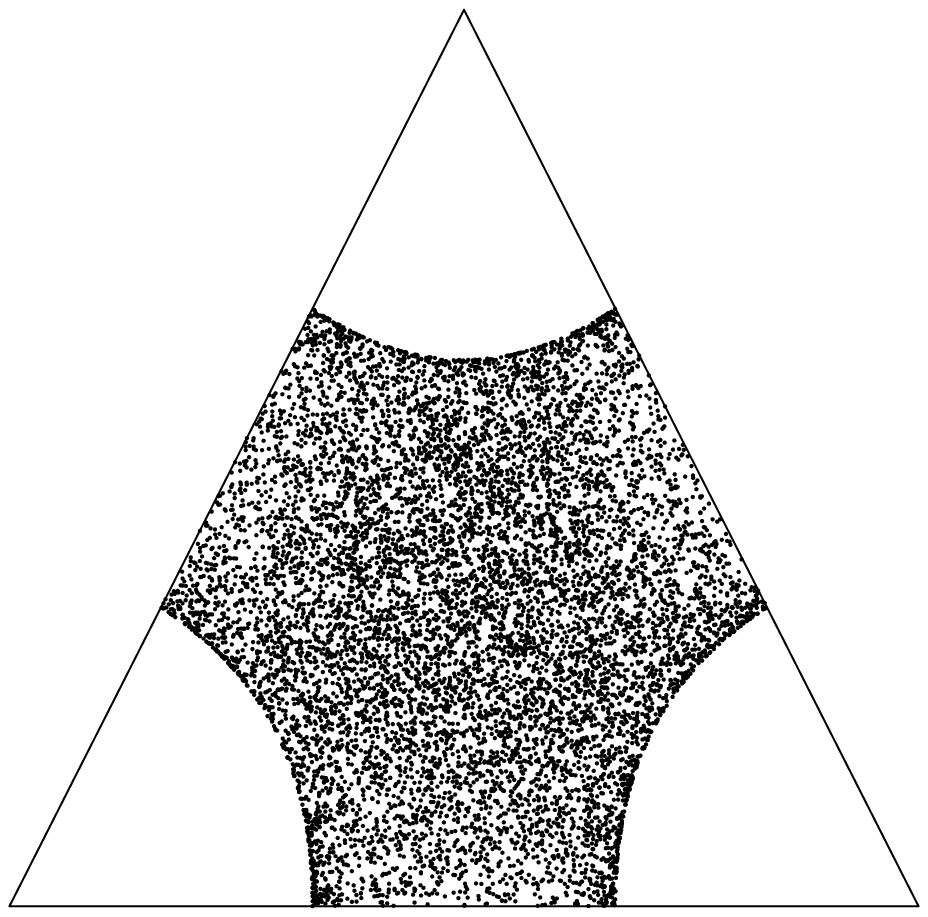}}\caption{Optimal strategies: classical
        (left) and quantum (right).}
         \label{qhex}
\end{figure}

Figure \ref{qhex} presents the areas (in both models) of frequency $q_m$ of appearance of individual
choice alternatives between two types of food, for  which  optimal strategies exist.
Let us observe that in the quantum case the area of the simplex corresponding to the optimal strategies has become slightly diminished in relation to the classical model.
The difference lies in the disappearance of areas at three boundaries of the regular hexagon which correspond to the arc-bounded surfaces.\footnote{As a curious detail, let us provide the precise size of such an area: $((\frac{1}{3}((9(-17+38\sqrt{2}\cdot3^{\frac{1}{4}}+10\sqrt{3}-22\sqrt{2}\cdot3^{\frac{3}{4}})+
(-939+2082\sqrt{2}\cdot3^{\frac{1}{4}}+542\sqrt{3}-1202\sqrt{2}\cdot3^\frac{3}{4})\pi)^{2}+(9(-72-8\sqrt{2}\cdot3^{\frac{1}{4}}+40\sqrt{3}+6\sqrt{2}\cdot3^{\frac{3}{4}})+(-3864-504\sqrt{2}\cdot3^{\frac{1}{4}}+2232\sqrt{3}+290\sqrt{2}\cdot3^{\frac{3}{4}}
 )\pi)^2))^\frac{1}{2})/(324((-3+\sqrt{2}\cdot3^{\frac{1}{4}}+2\sqrt{3}-\sqrt{2}
 \cdot3^{\frac{3}{4}})^2+(-2-2\sqrt{2}\cdot3^{\frac{1}{4}}+2\sqrt{3}+\sqrt{2}\cdot3^\frac{3}{4})^2))\approx$ 0.0120471.}
 Assuming the same measure of the possibility of occurrence of determined proportion of all three food pairs, we may say that the number of situations where the optimal strategies can be used in the quantum model makes up about 63$\%$ of all possibilities.
 In the classical variant, the area representing the optimal strategies makes up 67$\%$ of the simplex.
 This difference will be significant for analysis of intransitivity, which will be discussed more precisely in the next paragraph.
 It is also worth mentioning that in the classical model we deal with  sort of condensation of optimal strategies in the central part of the picture in the area of the balanced frequencies of all pairs of food. In the quantum case, they are more evenly spread, although they also appear less frequently  towards the sides of the triangle.

 \subsection{Intransitive orders}
 In the quantum model, we deal with an intransitive choice if one of the following conditions is fulfilled (\,see \cite{r1}):
 \begin{itemize}
  \item $P(C_2|B_1)=\frac{1-x_1}{2}<\frac{1}{2}$,
  $P(C_1|B_0)=\frac{1+x_2}{2}<\frac{1}{2}$,
  $P(C_0|B_2)=\frac{1-x_3}{2}<\frac{1}{2}$\,.
  \item $P(C_2|B_1)=\frac{1-x_1}{2}>\frac{1}{2}$,
  $P(C_1|B_0)=\frac{1+x_2}{2}>\frac{1}{2}$,
  $P(C_0|B_2)=\frac{1-x_3}{2}>\frac{1}{2}$\,.
 \end{itemize}
They form two  spherical equilateral triangles has tree equal $\frac{\pi}{2}$ angles.
 \begin{figure}[htbp]
          \centering{
        \includegraphics[
           height=2in,
           width=2.1in]%
          {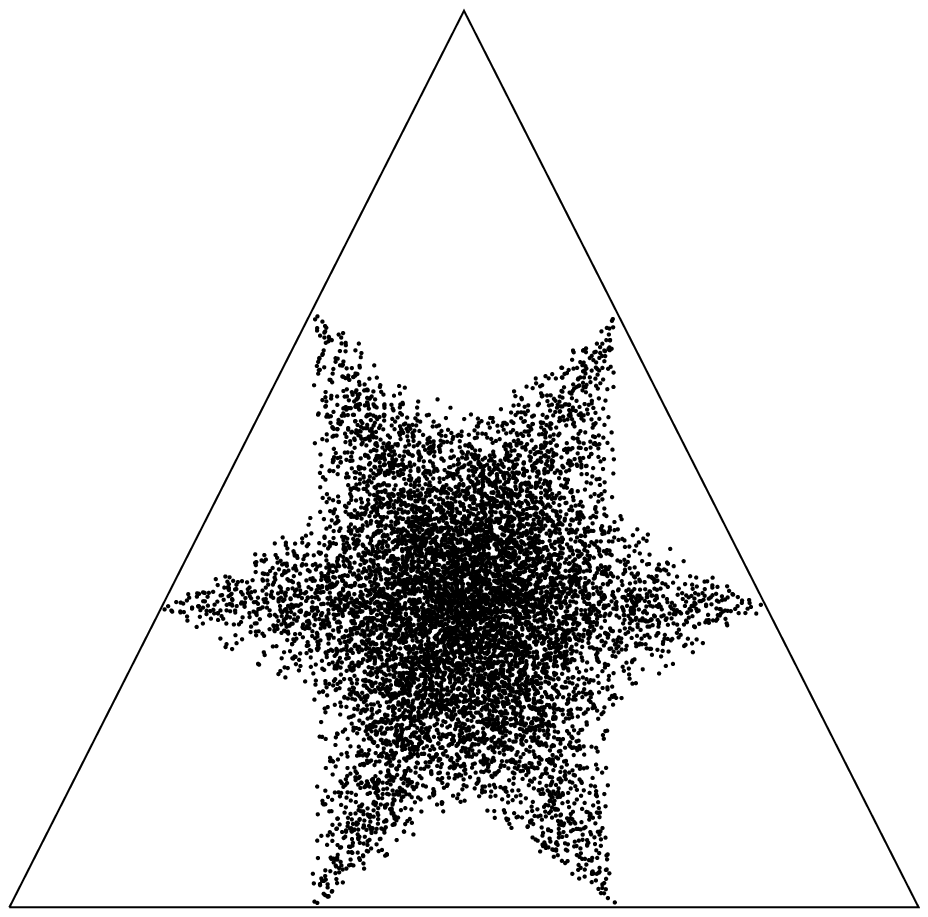}
          \includegraphics[
                       height=2in,
                       width=2.1
                       in]%
        {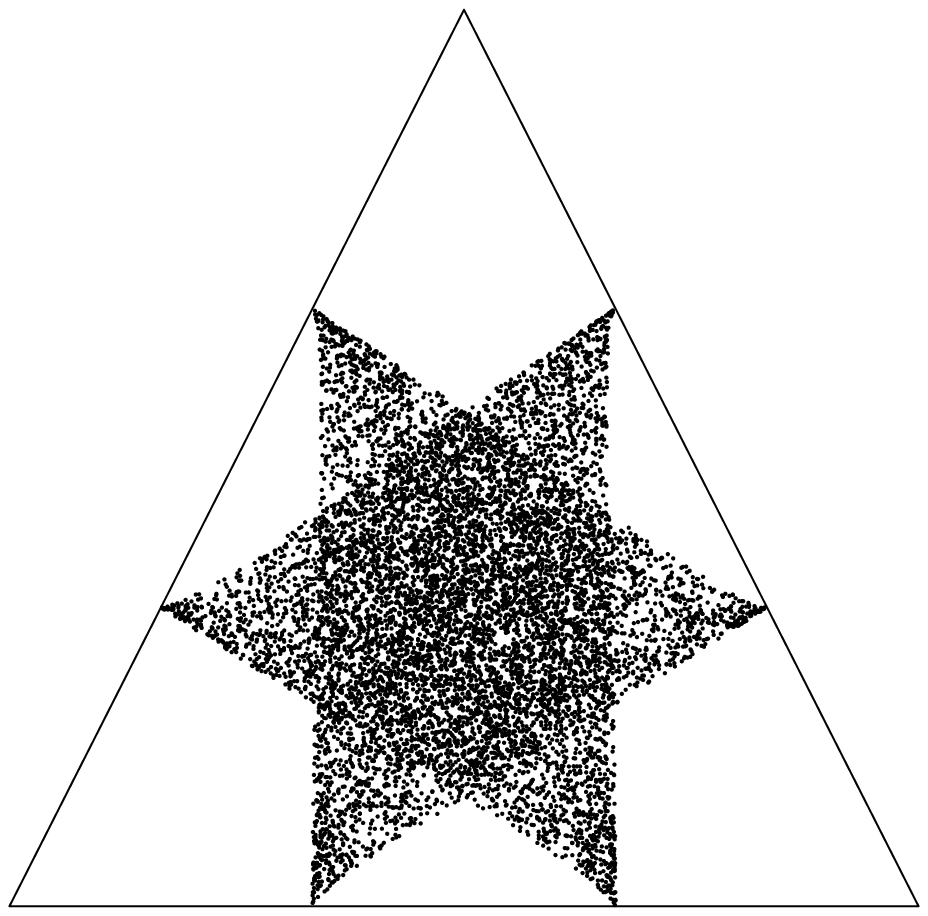}}\caption{Optimal intransitive strategies: classical
        (left) and quantum (right).}
          \label{qstar}
\end{figure}

It may be seen in Figure \ref{qstar} in what part of the simplex of parameters $(q_0,q_1,q_2)$ intransitive strategies may be used in both models. They form the six-armed star composed of two triangles\footnote{Any
of them corresponding to one of two possible intransitive orders.}
 in both the quantum and the classical model. As in the previous Figure one can notice that quantum variant is characterized by higher regularity, the star has clearly marked boundaries.
In both cases, we have got 33$\%$ \footnote {They are measured by the area of equilateral triangle inscribed
into a regular hexagon.}
 of conditions allowing to use intransitive optimal strategies in a determined order. There are 44$\%$ of conditions allowing to use intransitive strategies with an arbitrary order.
However, it is important to remember that in the quantum model, the number of all optimal strategies has decreased in relation to the classical variant. This, when the number of intransitive optimal strategies is equal, means that intransitive orders gain more importance in the quantum model. It is not the only reason leading to such a conclusion (see next paragraf).

\subsection{Transitive orders}

\begin{figure}[htbp]
          \centering{
        \includegraphics[
           height=2in,
           width=2.1in]%
          {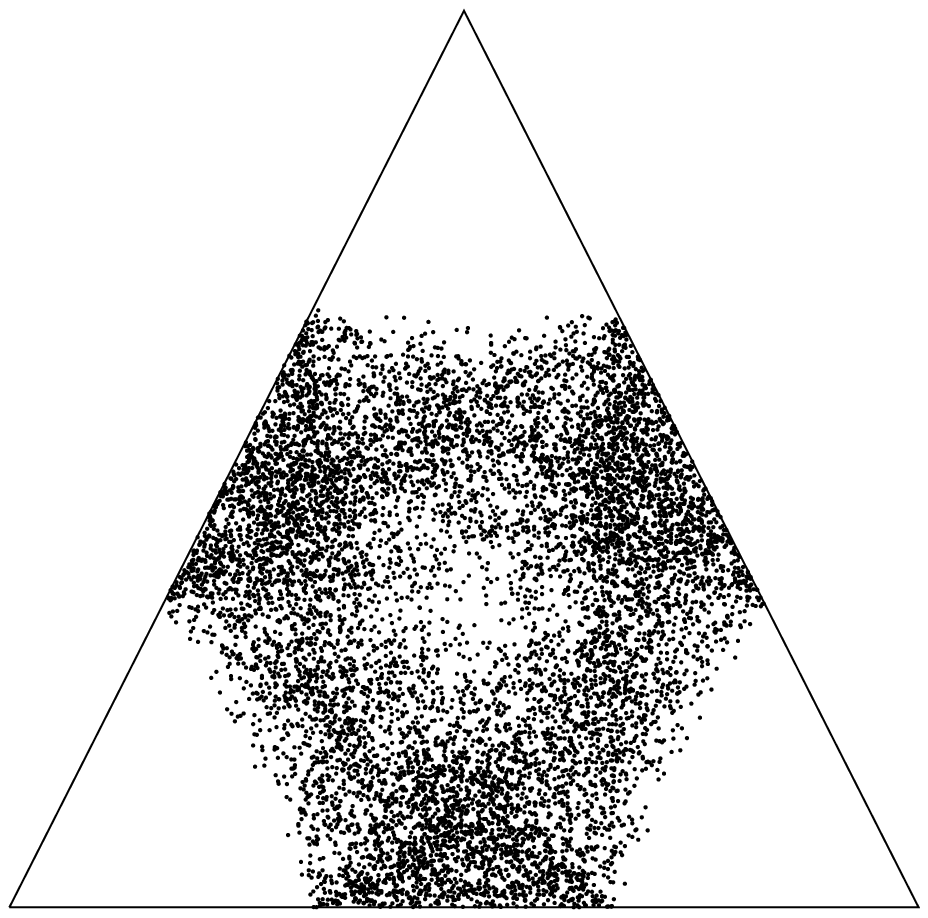}
          \includegraphics[
                       height=2in,
                       width=2.1
                       in]%
        {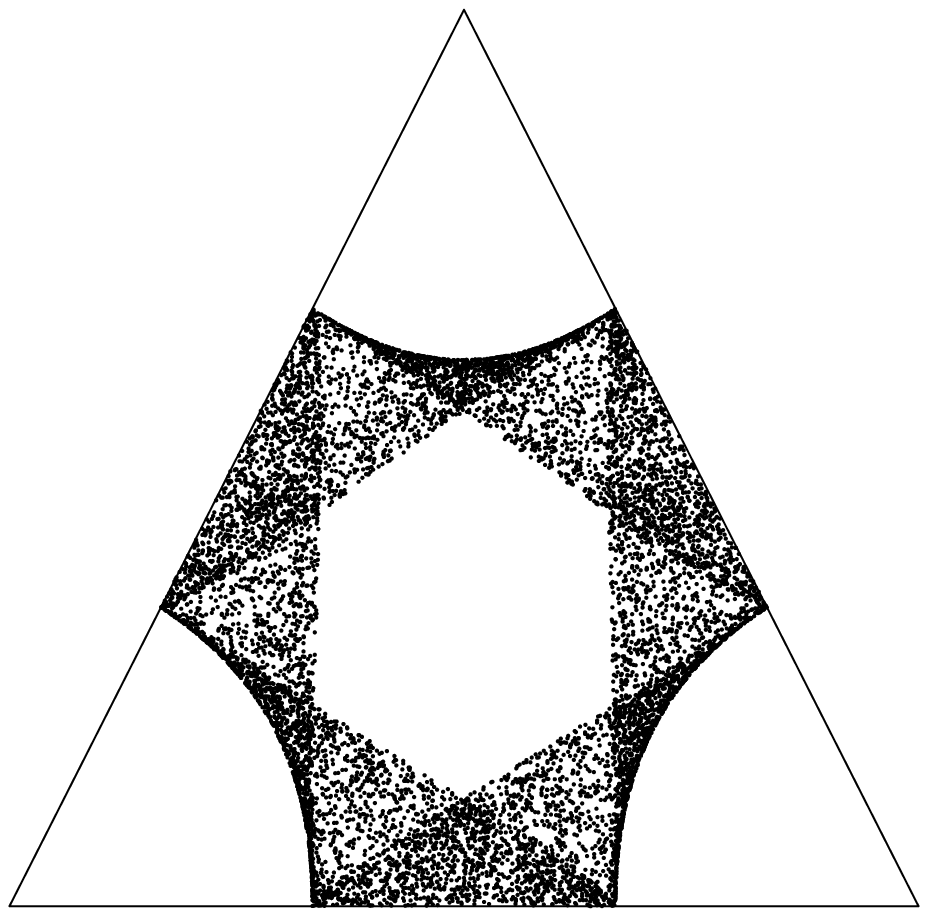}}\caption{Optimal transitive strategies: classical
        (left) and quantum (right).}
        \label{qtrans}
\end{figure}

Let us have a closer look at Figure \ref{qtrans}. It presents a simplex area for which there exist transitive optimal strategies in both models.
In the classical case optimal transitive strategies cover the same
area of the simplex as all optimal strategies, however they occur
less often in the center of the simplex (near point
$q_0=q_1=q_2=\frac{1}{3}$). The quantum version is essentially
different  - transitive optimal strategies do not appear
within the boundaries of a hexagon in the central part of the
picture (thus, there are about 41$\%$ of them). Let's observe that
this is the area where two different intransitive orders
superimpose (22$\%$).\footnote{The area of the regular six-armed
star is two times bigger than the area of the hexagon inscribed
into it.} Therefore, there cannot be defined for each intransitive
order a transitive order whose working effects are identical.
Moreover, the transitive strategies appear much less frequently
within the arms of the star forming intransitive orders. The above
remarks point to the fact that in the quantum model (within the
boundaries of the pure strategy) intransitive preferences
significantly gain more importance. To make the analysis clear,
let us sum up our quantitative discussion  by gathering the
results round into a table:
\begin{table}[htbp]
\caption{Comparison of achievability of various types of optimal
strategies in both models. } \vspace{1ex} \centering\footnotesize
\begin{tabular}{|c|c|c|c|} \hline
\vphantom{$\int^1$}&
 All &  Intransitive &  Transitive
\\[1pt]\hline
\vphantom{$\int^1$}Classical model \vphantom{$F^K$} & 67 $\%$ & 44 $\%$ & 67 $\%$
\\[1pt]\hline \vphantom{$\int^1$} Quantum model\vphantom{$F^K$}
 & 63 $\%$  & 44 $\%$ & 41 $\%$ \\[1pt]\hline
\end{tabular}
\end{table}
\subsection{Remark about quantum mixed strategies}
Any quantum cat's mixed strategy $\rho$
can be identified with a point $p$ inside a ball whose boundary is a set of pure strategies represented by a Bloch sphere  $S_2$.
A line passing through a point  $p$ and the centre of the ball cuts the sphere in two antipodal points $-\vec{v}$ and $\vec{v}$.
The point  $p$ divides the segment [$-\vec{v}$,\,$\vec{v}$] in the same ratio as the ratio of weights  $w_{v}$ i $w_{-v}$
in the representation of a mixed strategy $\rho$ as a convex combination of two pure strategies:
\begin{displaymath}
\rho = w_v |z_v\rangle\langle z_v|+ w_{-v} |z_{-v}\rangle\langle z_{-v}|.
\end{displaymath}
 Two antipodal points  $-\vec{v}$ i $\vec{v}$ of the sphere represent pure cat's strategies with the same property
   ( intransitive or transitive).\footnote{If coordinates of any vector $\vec{v}$  satisfy one of the conditions of intransitivity (see paragraph 5.2), then coordinates of $-\vec{v}$ satisfy the other one.}
  Since formulas  (\ref{prop}) are linear, each point lying on the segment [$-\vec{v}$,\,$\vec{v}$] will represent an strategy of the same property as points being the ends of this line. \newline
 The  randomized model  in which player operates mixed
strategies  has the unique property --- preferences of mixed
strategies are not different from preferences of respective pure
strategies lying on the line passing through the middle of the
sphere and a point inside the sphere specifying mixed strategy.

\section{Conclusion}
The aim of this work is to present some methods of
quantitative analysis of the, among others, intransitive orders
within the boundaries of the quantum game theory. We compared the results that the two models (quantum and classical) yield. The geometrical interpretation presented in
this article can turn out to be very helpful in
understanding  various quantum models in use.

It turns out that the order imposed by the player's rational
preferences can be intransitive. The quantum model gives a
considerable weight to intransitive orders. They are a constituent
part of more of all optimal strategies than in the classical case.
Moreover, for some frequencies of appearance of pairs of food,
quantum cat is able to achieve optimal results only thanks to the
intransitive strategy. It is a significant difference in reference
to classical cat's situation. However, it must be admitted here
that it refers only to the simple patterns of cat's behavior.
Perhaps, more advanced research  into  quantum game theory
will confirm validity of the intransitive decision algorithms,
which are often in contradiction with our intuition. Thorough
analysis of this problem would  be of great importance to those
who investigate our mind performance or for the construction of
thinking machines. 

 Mathematics have often been inspired by games. This gave rise
to the new fields of research ( studies of games of chance gave
rise to a large branch of mathematic called probability theory).
In our everyday lives, we encounter  various situations of
conflict and cooperation where we have to make particular
decisions. Many problems in the fields of economy and political
sciences  can  be expressed in the language of the quantum
game theory. In physics, the problem of measurement can be
considered as a game against Nature  - the observer  tries to
gain most possible information about the observed object. Other
experiments can be modelled in the same way.  Due to the
wide range of possible applications,  quantum game theory may shed
new light on the contemporary physics \cite{r29}. It may also
considerably influence the development of science and should
prepare us for the incoming  era of quantum computers. Therefore,
it is vital to carry on  research into this new field.

 \begin{center}
{\bf Acknowledgements}
\end{center}
This paper has been supported
by the {\bf Polish Ministry of Scientific Research and Information
Technology} under the (solicited) grant No {\bf
PBZ-MIN-008/P03/2003}.


\end{document}